\documentclass[12pt]{article}
\usepackage{lscape}
\usepackage{citesort}
\usepackage{amssymb}
\usepackage{appendix}
\usepackage{multirow}

\begin{document}
\def\qq{\langle \bar q q \rangle}
\def\uu{\langle \bar u u \rangle}
\def\dd{\langle \bar d d \rangle}
\def\sp{\langle \bar s s \rangle}
\def\GG{\langle g_s^2 G^2 \rangle}
\def\Tr{\mbox{Tr}}
\def\figt#1#2#3{
        \begin{figure}
        $\left. \right.$
        \vspace*{-2cm}
        \begin{center}
        \includegraphics[width=10cm]{#1}
        \end{center}
        \vspace*{-0.2cm}
        \caption{#3}
        \label{#2}
        \end{figure}
	}
	
\def\figb#1#2#3{
        \begin{figure}
        $\left. \right.$
        \vspace*{-1cm}
        \begin{center}
        \includegraphics[width=10cm]{#1}
        \end{center}
        \vspace*{-0.2cm}
        \caption{#3}
        \label{#2}
        \end{figure}
                }

\def\ds{\displaystyle}
\def\beq{\begin{equation}}
\def\eeq{\end{equation}}
\def\bea{\begin{eqnarray}}
\def\eea{\end{eqnarray}}
\def\beeq{\begin{eqnarray}}
\def\eeeq{\end{eqnarray}}
\def\ve{\vert}
\def\vel{\left|}
\def\ver{\right|}
\def\nnb{\nonumber}
\def\ga{\left(}
\def\dr{\right)}
\def\aga{\left\{}
\def\adr{\right\}}
\def\lla{\left<}
\def\rra{\right>}
\def\rar{\rightarrow}
\def\lrar{\leftrightarrow}  
\def\nnb{\nonumber}
\def\la{\langle}
\def\ra{\rangle}
\def\ba{\begin{array}}
\def\ea{\end{array}}
\def\tr{\mbox{Tr}}
\def\ssp{{\Sigma^{*+}}}
\def\sso{{\Sigma^{*0}}}
\def\ssm{{\Sigma^{*-}}}
\def\xis0{{\Xi^{*0}}}
\def\xism{{\Xi^{*-}}}
\def\qs{\la \bar s s \ra}
\def\qu{\la \bar u u \ra}
\def\qd{\la \bar d d \ra}
\def\qq{\la \bar q q \ra}
\def\gGgG{\la g^2 G^2 \ra}
\def\GG{\langle g_s^2 G^2 \rangle}
\def\g5{\gamma_5 \not\!q}
\def\x{\gamma_5 \not\!x}
\def\g5{\gamma_5}
\def\sb{S_Q^{cf}}
\def\sd{S_d^{be}}
\def\su{S_u^{ad}}
\def\sbp{{S}_Q^{'cf}}
\def\sdp{{S}_d^{'be}}
\def\sup{{S}_u^{'ad}}
\def\ssp{{S}_s^{'??}}

\def\sig{\sigma_{\mu \nu} \gamma_5 p^\mu q^\nu}
\def\fo{f_0(\frac{s_0}{M^2})}
\def\ffi{f_1(\frac{s_0}{M^2})}
\def\fii{f_2(\frac{s_0}{M^2})}
\def\O{{\cal O}}
\def\sl{{\Sigma^0 \Lambda}}
\def\es{\!\!\! &=& \!\!\!}
\def\ap{\!\!\! &\approx& \!\!\!}
\def\ar{&+& \!\!\!}
\def\arrr{\!\!\!\! &+& \!\!\!}
\def\ek{&-& \!\!\!}
\def\kek{\!\!\!\!&-& \!\!\!}
\def\cp{&\times& \!\!\!}
\def\se{\!\!\! &\simeq& \!\!\!}
\def\eqv{&\equiv& \!\!\!}
\def\kpm{&\pm& \!\!\!}
\def\kmp{&\mp& \!\!\!}
\def\mcdot{\!\cdot\!}
\def\erar{&\rightarrow&}


\def\simlt{\stackrel{<}{{}_\sim}}
\def\simgt{\stackrel{>}{{}_\sim}}


\title{
         {\Large
                 {\bf
Magnetic moments of $\Xi_Q^\prime$-$\Xi_Q$ transitions in light cone QCD
                 }
         }
      }

\author{\vspace{1cm}\\
{\small  T. M. Aliev$^1$ \thanks {taliev@metu.edu.tr}~\footnote{Permanent
address: Institute of Physics, Baku, Azerbaijan.}}\,\,
{\small  K. Azizi$^2$ \thanks {kazizi@dogus.edu.tr}}\,\,,
{\small M. Savc{\i}$^1$ \thanks
{savci@metu.edu.tr}} \\
{\small $^1$ Department of Physics, Middle East Technical University,
06531 Ankara, Turkey}\\
{\small $^2$ Department of Physics, Do\u gu\c s University,
Ac{\i}badem-Kad{\i}k\"oy, 34722 \.{I}stanbul, Turkey}}

\date{}

\begin{titlepage}
\maketitle
\thispagestyle{empty}

\begin{abstract}

The $\Xi_Q^\prime$-$\Xi_Q$ transition magnetic moments are calculated in
framework of the light cone QCD sum rules method (LCSR). The values of the
transition magnetic moments obtained are compared with the predictions of
the other theoretical approaches.

\end{abstract}

~~~PACS numbers: 11.55.Hx, 13.25.Hw, 13.25.Jx

\end{titlepage}

\section{Introduction}

During the last decade the heavy hadron spectroscopy has been a field of
permanent and growing interest. This interest dictated by the exciting
experimental results obtained in regard to this subject. At present all
ground state baryons containing a single charm quark have already been
observed and their masses are also measured. Many of the spin-1/2 hadrons
with a single bottom quark, such as $\Lambda_b$, $\Sigma_b$, $\Xi_b$ and
$\Omega_b$, and spin-3/2 hadrons such as $\Sigma_b^\ast$ have also been
observed \cite{Rtrn01,Rtrn02,Rtrn03}. Recently, spin-3/2 $\Xi_b^\ast$
baryon has been discovered \cite{Rtrn04}, and latest measurement of
the life-time of $\Lambda_b$ baryon has been announced \cite{Rtrn05}.
On the other hand the experimental situation with two and three heavy quarks
is not reach yet. Only SELEX collaboration announced the observation of the
spin-1/2 $\Xi_{cc}$ baryon \cite{Rtrn06}, which is not confirmed yet by the
other collaborations.

Experimental observation and investigation of the properties of doubly and
triply heavy baryons constitute one of the most promising research area in
particle physics.
One of the main static quantities which could give
valuable information about the internal structure of baryons is their magnetic
moments. Magnetic moment of heavy baryons have extensively been investigated
within the framework of different approaches like, naive quark model, chiral
quark model, non-relativistic quark model and QCD sum rules method.

The main advantage of the QCD sum rules method compared to the other
non-perturbative approaches is that, it is based on the fundamental QCD
Lagrangian, and it takes into account the non-perturbative nature of the QCD
vacuum. In order to solve the problems which are inherit in the traditional
QCD sum rules method, the light cone version of the QCD sum rules is
proposed (for more about this method, see for example \cite{Rtrn07}).
In this version of the QCD sum rules, the operator product expansion
is performed over twists of the operators and non-perturbative effects are
all encoded into the matrix elements of the non-local operators between
the vacuum and one-particle states. The light cone QCD sum rules (LCSR) has
so far been applied to many problems of the hadron physics (for the very
recent applications of the LCSR, see for example
\cite{Rtrn08,Rtrn09,Rtrn10}).

In the present work we calculate the $\Xi_Q^\prime$-$\Xi_Q$ transition
magnetic moment within the QCD sum rules method. Note that
the $\Lambda_Q$-$\Sigma_Q$ transition magnetic moment was studied within
the same framework in \cite{Rtrn11}.  

The paper is organized as follows. In section 2 the sum rules for the
$\Xi_Q^\prime$-$\Xi_Q$ transition magnetic moment are obtained. The
following section contains numerical calculations, discussion and comparison
with the predictions of the other theoretical methods existing in
literature. 

\section{Derivation of the sum rules for the $\Xi_Q^\prime$-$\Xi_Q$
transition magnetic moments in LCSR}

We start this section by summarizing the useful information on $SU(3)$
classification of the heavy hadrons with single heavy quarks. These baryons
belong to either  $SU(3)$ antisymmetric $\bar{3}_F$ or symmetric $6_F$
flavor representations. It is well known that, in $6_F$ representation the
ground state  baryons must have total spin 1, while their total spin is 0
in $\bar{3}_F$ representation. Therefore, baryons in $6_F$ representation
can have spin-1/2 or -3/2, but hadrons in $\bar{3}_F$ representation can have
only spin-1/2. In the present work we consider only spin-1/2 baryons
from both representations, i.e., we consider $\Xi_Q^\prime$ from $6_F$ and 
$\Xi_Q$ from $\bar{3}_F$ representations, respectively.

Following this brief information, we can proceed now to calculate the
$\Xi_Q^\prime$-$\Xi_Q$ transition magnetic moment within LCSR. For this aim
we consider the following correlation function:
\bea
\label{etrn01}
\Pi = i\int d^x e^{ipx} \lla 0 \vel \mbox{T} \Big\{ \eta_{\Xi_Q} (x)
\bar{\eta}_{\Xi_Q^\prime} (0) \Big\} \ver 0 \rra_\gamma~,
\eea
where $\gamma$ is the external electromagnetic field; $\eta_{\Xi_Q}$ and
$\eta_{\Xi_Q^\prime}$ are the interpolating currents of $\Xi_Q$ and
$\Xi_Q^\prime$ baryons. The general forms of the interpolating currents of
$\Xi_Q$ and $\Xi_Q^\prime$ baryons are as follows:
\bea
\label{etrn02}
\eta_{\Xi_Q} \es {1 \over \sqrt{6}} \epsilon^{abc} \left\{ 2 \left( s^{aT} C
q^b \right) \gamma_5 Q^c + \left( s^{aT} C Q^b \right) \gamma_5 q^c -  
\left( q^{aT} C Q^b \right) \gamma_5 s^c \right.\nnb \\
\ar \left.2 t \left( s^{aT} C \gamma_5 q^b \right) Q^c + t \left( s^{aT} C \gamma_5 Q^b
\right) q^c - t \left( q^{aT} C \gamma_5 Q^b \right) s^c \right\}~,\nnb \\ \nnb \\
\eta_{\Xi_Q^\prime} \es {1\over \sqrt{2}} \epsilon^{abc} \left\{
\left( s^{aT} C Q^b \right) \gamma_5 q^c + \left( q^{aT} C Q^b \right) \gamma_5
s^c \right. \nnb \\
\ar  \left. t \left( s^{aT} C \gamma_5 Q^b \right) q^c + t \left( q^{aT} C
\gamma_5 Q^b \right) s^c \right\}~, 
\eea    
where $a,b,c$ are the color indices; $q=u$ or $d$, and $Q=b$ or $c$ quark;
$C$ is the charge conjugation operator; and $t$ is the arbitrary parameter.

It should be noted here that, $\Xi_Q$ and
$\Xi_Q^\prime$ baryons do have the same quantum numbers and in principle
there can be mixing between them. This mixing can be implemented by
modifying the interpolating currents for physical states as follows:
\bea
\label{etrn03}
\eta_{\Xi_Q}^{phys} \es \cos\theta_Q \, \eta_{\Xi_Q} + \sin\theta_Q \,
\eta_{\Xi_Q^\prime}~, \nnb \\
\eta_{\Xi_Q^\prime}^{phys} \es - \sin\theta_Q \, \eta_{\Xi_Q} + \cos\theta_Q
\, \eta_{\Xi_Q^\prime}~.
\eea
It is shown in \cite{Rtrn12} that the mixing angle between $\eta_{\Xi_b}
(\eta_{\Xi_c})$
and $\eta_{\Xi_b^\prime}(\eta_{\Xi_c^\prime})$ is equal to $\theta_b=6.4\pm 1.8^0$
($\theta_c=5.5\pm 1.8^0$), which is quite small, and hence we can safely
neglect it. According to sum rules method philosophy the correlation
function is calculated in terms of hadrons, and in terms of quarks and
hadrons in two different kinematical regions. Equating then the obtained
results to each other one can get the sum rules for the appropriate physical
quantity. 

We start calculating the correlation function from the hadronic side.
Inserting complete set of hadrons with $\eta_{\Xi_Q}$ and
$\eta_{\Xi_Q^\prime}$ state quantum numbers, and isolating the ground state
contributions, we get:
\bea
\label{etrn04}
\Pi \es {\lla 0 \vel \eta_{\Xi_Q} \ver \Xi_Q(p_1) \rra \over
(m_{\Xi_Q}^2-p_1^2)}
\lla \Xi_Q(p_1) \vel\right. \Xi_Q^\prime (p_2) \rra_\gamma
{\lla \Xi_Q^\prime (p_2) \vel \bar{\eta}_{\Xi_Q^\prime} \ver 0\rra  \over
(m_{\Xi_Q^\prime}^2-p_2^2)} + \cdots~,
\eea
where $p_2=p_1+q$, $p_1=p$ and $q$ is the photon momentum; and dots
represent the contributions of higher states and continuum. The matrix
element of the interpolating current between one baryon and vacuum states
is defined in the following way:
\bea
\label{etrn05}
\lla 0 \vel \eta_{\Xi_Q} \ver \Xi_Q(p) \rra \es \lambda_{\Xi_Q} u_{\Xi_Q}(p)~, \nnb \\
\lla \Xi_Q^\prime (p)\vel \bar{\eta}_{\Xi_Q^\prime} \ver 0\rra \es
\lambda_{\Xi_Q^\prime} \bar{u}_{\Xi_Q^\prime}(p)~.
\eea
The second matrix element in (\ref{etrn04}) is parametrized in terms of two
form factors $F_1$ and $f_2$ as:
\bea
\label{etrn06}
\lla \Xi_Q(p_1) \vel\right. \Xi_Q^\prime (p_2) \rra_\gamma \es
\bar{u}(p_1) \left[f_1 \gamma_\mu - {i \sigma_{\mu\nu}q^\nu \over
m_{\Xi_Q} + m_{\Xi_Q^\prime}} f_2 \right] u(p_2) \varepsilon^\mu~,\nnb \\
\es \bar{u}(p_1) \left[ (f_1+f_2) \gamma_\mu - f_2 (p_1+p_2)_\mu \right]
u(p_2) \varepsilon^\mu~,
\eea
where we set $q^2=0$ for the real photon. Using Eqs. (\ref{etrn05}) and
(\ref{etrn06}) in hadronic side of the correlation function, we get the
following expression:
\bea
\label{etrn07}
\Pi = {\lambda_{\Xi_Q} \lambda_{\Xi_Q^\prime} \over m_{\Xi_Q}^2 - p_1^2}
{(\rlap/{p}_1+m_{\Xi_Q})\over m_{\Xi_Q^\prime}^2-p_2^2} \left\{(f_1+f_2)
\gamma_\mu - f_2 (p_1+p_2)_\mu \right\} (\rlap/{p}_2+m_{\Xi_Q^\prime})\varepsilon^\mu~.
\eea
We see from this expression that there are many structures appearing in the
phenomenological part of the correlation function in calculation of the
transition magnetic moment. In the present work we chose the structure
$\rlap/{p}_1 \rlap/{\varepsilon} \rlap/{p}_2$ in determining the transition
magnetic form factor $f_1+f_2$. At $q^2=0$ this combination gives the transition
magnetic moments in natural units, .i.e.,
$e\hbar/(m_{\Xi_Q}+m_{\Xi_Q^\prime})$. The coefficient of the structure $\rlap/{p}_1
\rlap/{\varepsilon} \rlap/{p}_2$ gives the expression of the invariant
function for the transition magnetic moment $\mu_{\Xi_Q \Xi_Q^\prime}$ in
the following form:
\bea
\label{etrn08}
\Pi = {\lambda_{\Xi_Q} \lambda_{\Xi_Q^\prime} \over
\left(m_{\Xi_Q}^2 - p_1^2\right) \left(m_{\Xi_Q^\prime}^2 - p_2^2 \right) }
\mu_{\Xi_Q \Xi_Q^\prime}~.
\eea

Theoretical part of the correlation function is calculated in the following
way. At the first step the correlation function is written in terms the
quark operators using the Wick theorem. The correlation function contains
two parts, namely, photon interacting with quarks perturbatively, and
photon interacting with quarks non-perturbatively. In order to calculate the
perturbative contribution it is enough replacing one of the three
propagators with
\bea
\label{etrn09}
S \to -{1\over 2} \int dy\, S^{free} (x-y) \gamma_\mu S^{free} (y) y_\nu {\cal
F}_{\mu\nu}~,
\eea
and the remaining two quark propagators are taken as free quark
propagators. In Eq. (\ref{etrn09}) 
the fixed point gauge (the so-called Fock-Schwinger gauge) is
used, i.e., $A_\mu = (1/2) y_\nu {\cal F}_{\mu\nu}$, and $S^{free}$ is the
free quark operator. Free quark operator for the light and
heavy quarks is given as:
\bea
\label{etrn10}
S^{free}(x) \es {i \rlap/x\over 2\pi^2 x^4} - {m_q\over 4 \pi^2 x^2} \nnb \\
S_Q^{free}(x) \es {m_Q^2 \over 4 \pi^2} {K_1(m_Q\sqrt{-x^2}) \over \sqrt{-x^2}} -
i {m_Q^2 \rlap/{x} \over 4 \pi^2 x^2} K_2(m_Q\sqrt{-x^2})~,
\eea
respectively, where
$K_i$ are the modified Bessel function of the second kind; and $m_q$ and
$m_Q$ are the masses of the light and heavy quarks, respectively. Following the
above-explained steps, the perturbative part of the correlation function
can be calculated straightforwardly.

The non-perturbative contribution to the correlation function can be
calculated by replacing one of the light quark propagators with
\bea
\label{etrn11}
S_{\alpha\beta}^{ab} = - {1\over 4} \bar{q}^a \Gamma_i q^b
\left(\Gamma_i\right)_{\alpha\beta}~,
\eea
where $\Gamma_i$ is the full set of the Dirac matrices, i.e., $\Gamma_i=
\left\{ I,\gamma_5,\gamma_\mu,i\gamma_5\gamma_\mu,\sigma_{\mu\nu}/\sqrt{2}
\right\}$; and the remaining two propagators are taken as full quark
propagators. So, for calculation of the perturbative and non-perturbative
parts of the correlation function we need the expressions of the light and
heavy propagators in external field.

The light cone expansion of the propagator in external field is performed in
\cite{Rtrn13} and it receives contributions from non-local four-quark
$\bar{q}q\bar{q}q$, three-particle $\bar{q}G q$, four-particle $\bar{q}G
G q$, etc. operators, where $G$ is the gluon field strength tensor. In the
present work we take into account contributions coming only from non-local
operators with one gluon. The contributions of four-quark and
two-gluon-two-quark operators are neglected due to their small
contributions. Under these approximations the heavy and light quark
operators, in presence of the external field, have the following forms:
\bea
\label{etrn12}
i S_{light}(x) \es i S_{light}^{free}(x) \nnb \\
\ek {\lla \bar q q \rra\over 12} \left(1 - {m_0^2 x^2 \over 16 }\right)
- i g_s \int_0^1 du \left[{\rlap/x\over 16 \pi^2 x^2} G_{\mu \nu} (ux)
\sigma_{\mu \nu} - {i\over 4 \pi^2 x^2} u x^\mu G_{\mu \nu} (ux) \gamma^\nu
\right]~, \nnb \\ \nnb \\
i S_{heavy}(x) \es i S_{heavy}^{free}(x) \nnb \\
\ek ig_s \int {d^4k \over (2\pi)^4} e^{-ikx} \int_0^1
du \Bigg[ {\rlap/k+m_Q \over 2 (m_Q^2-k^2)^2} G^{\mu\nu} (ux)
\sigma_{\mu\nu} +
{u \over m_Q^2-k^2} x_\mu G^{\mu\nu} \gamma_\nu \Bigg]~.
\eea    
It follows from Eqs. (\ref{etrn11}) and (\ref{etrn12}) that, in order 
to calculate the non-perturbative contribution to the correlation function we
need to know the following matrix elements
$\lla \gamma(q) \vel \bar{q} \Gamma_i q \ver 0 \rra$ and
$\lla \gamma(q) \vel \bar{q} \Gamma_i G_{\mu\nu} q \ver 0 \rra$. These
matrix elements are defined in terms of the photon distribution amplitudes
(DAs), whose expressions all can be found in \cite{Rtrn14}.
Using Eq. (\ref{etrn12}) and
definitions of the photon DAs one can calculate the theoretical part of the
correlation function. The sum rules for the magnetic moment of
$\Xi_Q^\prime$-$\Xi_Q$ transition is obtained by equating the coefficients
of the structure $\rlap/{p}_1 \rlap/{\varepsilon} \rlap/{p}_2$
from both sides of the correlation function. The final step in deriving the
sum rules for the magnetic moment of the $\Xi_Q^\prime$-$\Xi_Q$ transition
is performing double Borel transformation over the variables $-p_1^2 \to
M_1^2$ and $-p_2^2 = (p_1+q)^2 \to M_2^2$ in order to suppress higher state
and continuum contributions. Finally we get:
\bea
\label{etrn13}
\lambda_{\Xi_Q} \lambda_{\Xi_Q^\prime} \mu_{\Xi_Q \Xi_Q^\prime} = 
e^{\left(m_{\Xi_Q}^2/M_1^2 + m_{\Xi_Q^\prime}^2/M_2^2\right)} \Pi^{theor}~.
\eea
The expression for $\Pi^{theor}$ is rather lengthy, therefore we do
not present it here.

It should be noted here that, since the masses of $\Xi_Q$ and $\Xi_Q^\prime$
are very close to each other, we can set $M_1^2=M_2^2$ and $m_{\Xi_Q}=
m_{\Xi_Q^\prime}$.

It follows from Eq. (\ref{etrn13}) that in calculation of the transition
magnetic moment $\mu_{\Xi_Q \Xi_Q^\prime}$ we need to know the residues of
the $\Xi_Q$ and $\Xi_Q^\prime$ baryons. The residues of sextet and
anti-triplet heavy baryons have already been calculated in \cite{Rtrn15},  
and for this reason we do not present them in this work.

\section{Numerical analysis}

In this section we perform numerical analysis of the sum rules for the
transition magnetic moment obtained in the previous section. The main input
parameters of the sum rules are photon DAs, whose explicit expressions are
given \cite{Rtrn14}.

The remaining input parameters which we use in the numerical analysis of the
sum rules are: $\qq~(1~GeV)=-(0.243)^3~GeV^3$, $\qs~(1~GeV)=0.8\qq~(1~GeV)$,
$m_0^2 = (0.8\pm0.2)~GeV^2$ \cite{Rtrn16}, the magnetic
susceptibility $\chi~(1~GeV)=-2.85 \pm0.5~GeV^{-2}$ \cite{Rtrn17}.
The sum rules contain the following auxiliary parameters: Borel mass
square $M^2$, continuum threshold $s_0$, and the arbitrary parameter
$\beta$ in the interpolating current. Obviously one expects the transition
magnetic moment be independent of these parameters. For this reason in
further numerical analysis we shall look for the 
regions of these parameters where magnetic moment is practically independent
of them. The domain of $M^2$ is determined from the following two
conditions:

(i) The ground state contribution should be larger compared to higher
states and the continuum contributions

(ii) the highest terms in $1/M^2$ should constitute, say about 30\% - 40\%
of the higher power $M^2$ terms.

Our numerical analysis shows that both
conditions are satisfied when $10 \le M^2 \le 25~GeV^2$ for the $\Xi_b$
baryon, and $6 \le M^2 \le 12~GeV^2$ for the $\Xi_c$ baryon, respectively.
The working regions of $s_0$ and $\beta$ are also determined from the
consideration that the magnetic moment should not change appreciably in the
respective regions of these parameters.

The dependence of the magnetic moments for the $\Xi_b^{\prime 0}$-$\Xi_b^0$,
$\Xi_b^{\prime -}$-$\Xi_b^-$, $\Xi_c^{\prime 0}$-$\Xi_c^0$ and
$\Xi_c^{\prime -}$-$\Xi_c^-$ transitions on $\cos\theta$, where
$\beta=\tan\theta$, at fixed values of $M^2$ and $s_0$ are presented in Figs.
(1)-(4), respectively. We see from these figures that the working region
for $\beta$ in the $b$-baryon case is $-0.5 \le \cos\theta \le -0.1$, and
for the baryons containing $c$-quark it is $-0.4 \le \cos\theta \le 0.1$.
We deduce from these figures the following values for the $\Xi_Q^{\prime
0}$-$\Xi_Q^0$ transition magnetic moments: $\mu_{\Xi_b^{\prime 0}\Xi_b^0} =
(1.4 \pm 0.1) \mu_N$, $\mu_{\Xi_b^{\prime -}\Xi_b^-} = (0.21 \pm 0.01) \mu_N$, 
$\mu_{\Xi_c^{\prime +}\Xi_c^+} = (1.3 \pm 0.1) \mu_N$ and
$\mu_{\Xi_c^{\prime 0}\Xi_c^0} = (0.18 \pm 0.02) \mu_N$, where $\mu_N$ is
the nuclear magneton. In the case of
$b$-baryons the change in the values of the magnetic moments is about 10\%
if $s_0$ varies in the domain $40 \le s_0 \le 45~GeV^2$. Hence we can safely be
say that the results for the magnetic moments are stable with respect to the
variation of $s_0$ in the determined domain. In Table 1 we present our
results on the transition magnetic moments together with the ones predicted
by different approaches, such as effective quark mass and screened quark
charge scheme (for $b$-hadron sector see \cite{Rtrn18}, 
for $c$-hadron sector see \cite{Rtrn19}),
bag model \cite{Rtrn20}, non-relativistic quark model
\cite{Rtrn21}, and chiral constituent quark model \cite{Rtrn22}.


\begin{table}[h]

\renewcommand{\arraystretch}{1.3}
\addtolength{\arraycolsep}{-0.5pt}
\small
$$
\begin{array}{|l|c|c|c|c|c|c|c|c|}
\hline \hline
\mbox{Transition}         & \mbox{Our result} &\cite{Rtrn18}
                                              &\cite{Rtrn18}  
                                              &\cite{Rtrn19}   
                                              &\cite{Rtrn19}
                          &\cite{Rtrn20} &\cite{Rtrn21}
                          &\cite{Rtrn22}                                                              \\ \hline
\Xi_b^{\prime 0}-\Xi_b^0  & 1.40 \pm 0.10     & 1.392 & 1.354 & \cdots     & \cdots     & 0.917 & 1.41  &   \cdots   \\ 
\Xi_b^{\prime -}-\Xi_b^-  & 0.21 \pm 0.01     & 0.178 & 0.142 & \cdots     & \cdots     & 0.82  & 0.16  &   \cdots   \\
\Xi_c^{\prime +}-\Xi_c^+  & 1.3  \pm 0.1      &  \cdots    &  \cdots    & -1.41 & -1.39 & 1.043 & 1.4   &  1.3  \\
\Xi_c^{\prime 0}-\Xi_c^0  & 0.18  \pm 0.02    &  \cdots    &  \cdots    & 0.18  & 0.13  & 0.013 & 0.08  & -0.31 \\
\hline \hline
\end{array}
$$
\caption{The $\Xi_Q^{\prime 0}-\Xi_Q^0$ transition magnetic moments
predicted by the light cone QCD sum
rules (our result), effective quark mass scheme \cite{Rtrn18,Rtrn19},
screened quark charge scheme \cite{Rtrn18,Rtrn19}, bag model \cite{Rtrn20},
non-relativistic quark model \cite{Rtrn21}, and
chiral constituent quark model  \cite{Rtrn22} approaches, in units of nuclear
magneton $\mu_N$. The first (second) column of \cite{Rtrn18} (\cite{Rtrn19})
corresponds the effective quark mass (screened quark charge) scheme.}
\renewcommand{\arraystretch}{1}
\addtolength{\arraycolsep}{-1.0pt}
\end{table}

From a comparison of our results with the predictions of the above-mentioned
approaches we see that for the $\Xi_c^{\prime +}-\Xi_c^+$ transition all
results are very close to each other. In the case of $\Xi_b^{\prime
0}-\Xi_b^0$ transition, except the bag model the results of all other approaches
are in good agreement. Our prediction for the $\Xi_b^{\prime -}-\Xi_b^-$  
transition is very close to the results of  \cite{Rtrn18} and
\cite{Rtrn21}, but considerably different from that of  \cite{Rtrn20}. As
far as $\Xi_c^{\prime 0}-\Xi_c^0$ transition is concerned, our result
confirms the prediction of \cite{Rtrn19}, it is approximately 1.5 times
smaller than that of \cite{Rtrn22}, two times larger than that of
\cite{Rtrn21} and 10 times larger than the results predicted in
\cite{Rtrn20}.

Finally we express our concluding remarks about the present work, in which
we calculate the magnetic moments of the $\Xi_Q^{\prime 0}-\Xi_Q^0$ ($Q=b$
or $c$) transitions within the framework of light cone QCD sum rules.
We also perform a comparison of our results with the
predictions of various approaches existing in literature. We observe that,
for the $ \Xi_b^{\prime 0}-\Xi_b^0$ and $ \Xi_c^{\prime +}-\Xi_c^+$ 
transitions the predictions of all approaches do practically agree with each
other. Similar situation takes place for the $\Xi_b^{\prime -}-\Xi_b^-$
transition, except the bag model prediction. As it comes to $\Xi_c^{\prime
0}-\Xi_c^0$ transition, our results coincide with the ones predicted in
\cite{Rtrn16}, but differ substantially with the predictions of all other
approaches. At present, these magnetic moments have not yet been measured in
experiments, and we hope that they all can be measured in future planned
experiments. We believe that the measurement of $\Xi_Q^{\prime 0}-\Xi_Q^0$
transition magnetic moments will also be very useful in determination of the
mixing angle among heavy baryons.

\newpage

\section*{Figure captions}
{\bf Fig. 1} The dependence of the transition magnetic moment
$\mu_{\Xi_b^{\prime 0}\Xi_b^0}$ on $\cos\theta$ at several fixed values of
the Borel mass-square $M^2$, and at fixed value of the 
continuum threshold $s_0=45~GeV^2$. \\ \\
{\bf Fig. 2} The same as in Fig. 1, but for the
$\mu_{\Xi_b^{\prime -}\Xi_b^-}$ transition. \\ \\
{\bf Fig. 3} The dependence of the transition magnetic moment
$\mu_{\Xi_c^{\prime +}\Xi_c^+}$ on $\cos\theta$ at several fixed values of
the Borel mass-square $M^2$, and at fixed value of the
continuum threshold $s_0=10~GeV^2$. \\ \\
{\bf Fig. 4} The same as in Fig. 1, but for the
$\mu_{\Xi_c^{\prime 0}\Xi_c^0}$ transition.

\newpage

\begin{figure}  
\vskip 3. cm
    \includegraphics{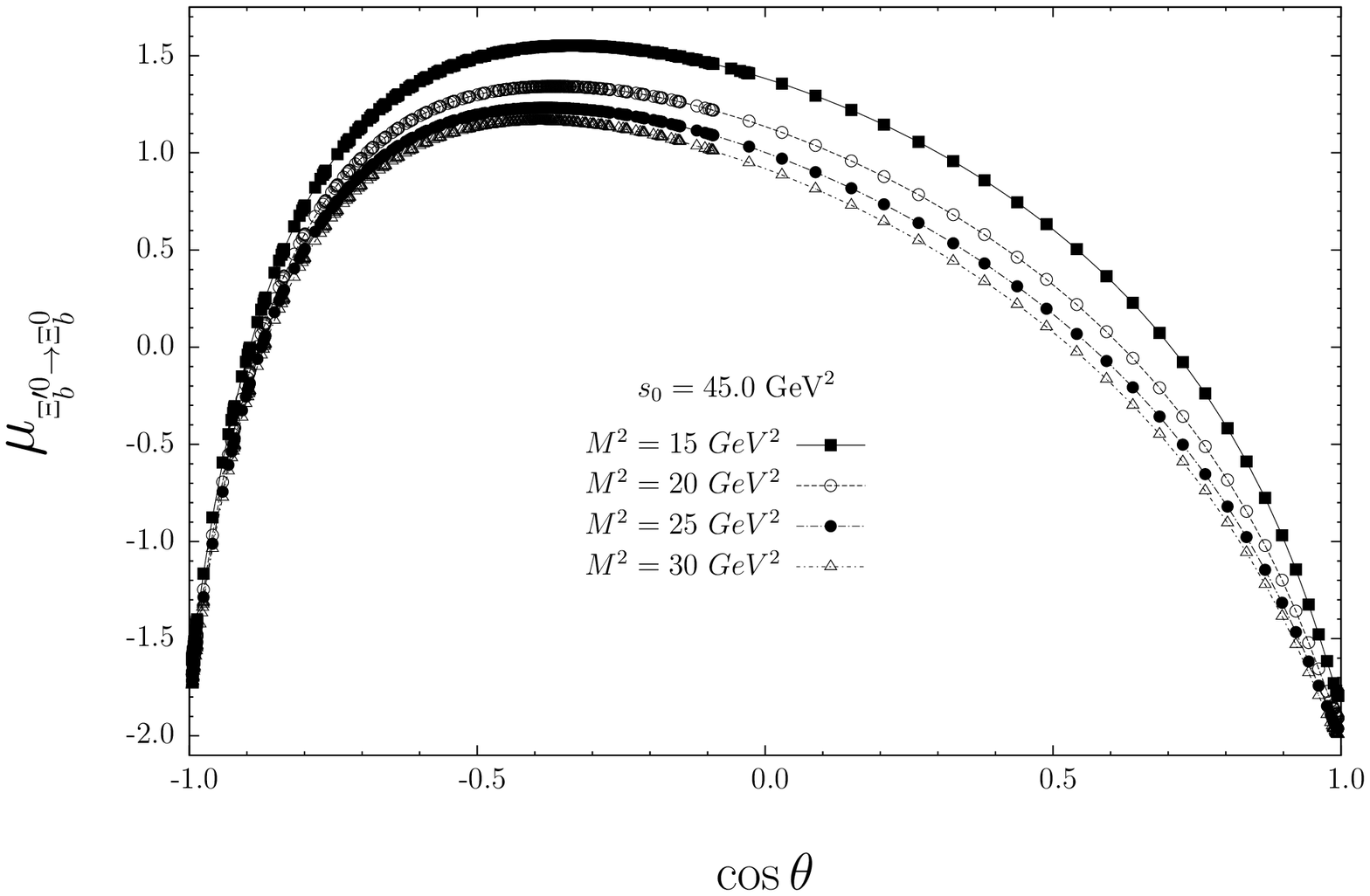}
\vskip 7.0cm   
\caption{}
\end{figure}

\begin{figure}
\vskip 3. cm
    \includegraphics{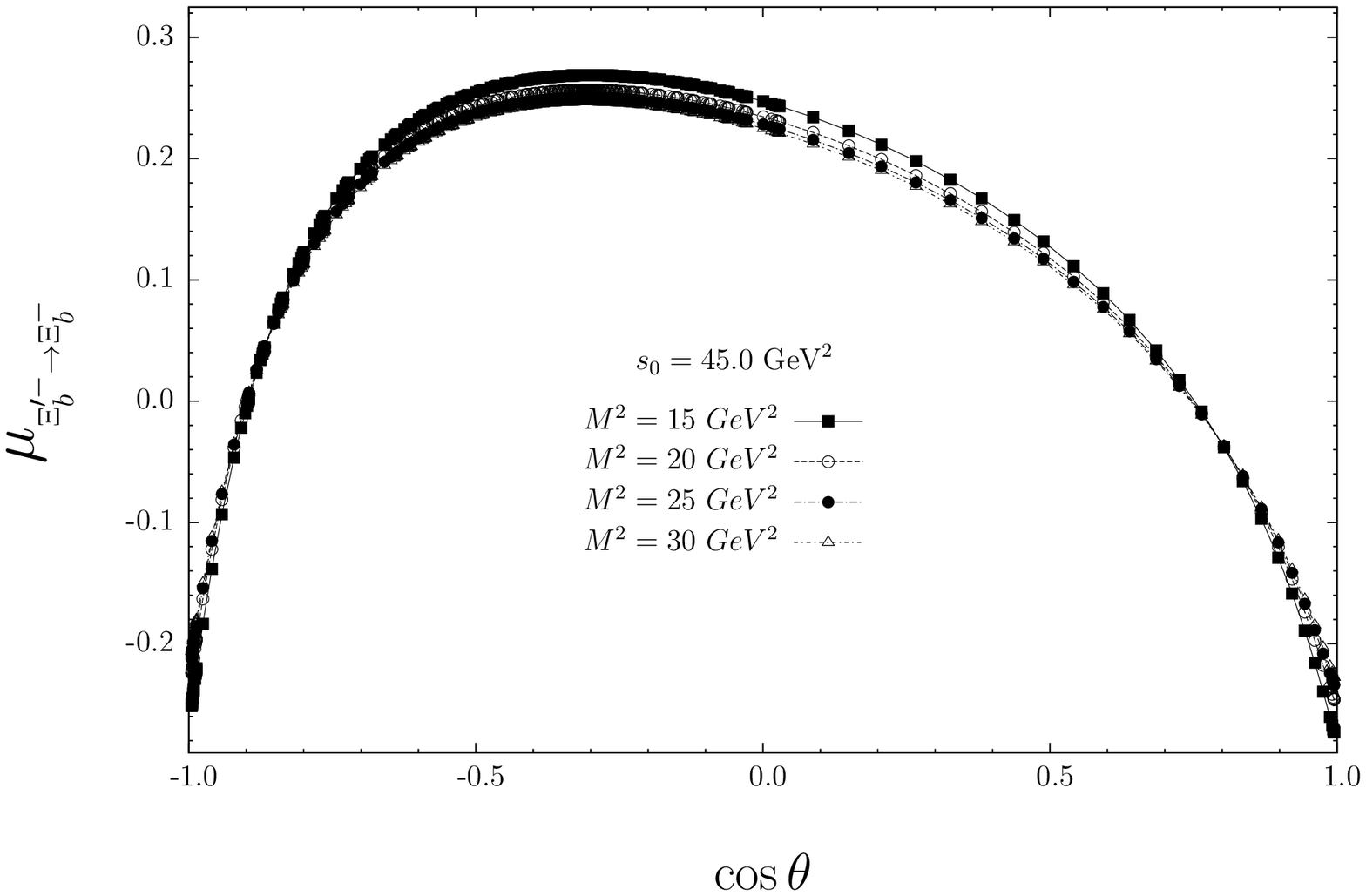}
\vskip 7.0cm
\caption{}
\end{figure}

\begin{figure}  
\vskip 3. cm
    \includegraphics{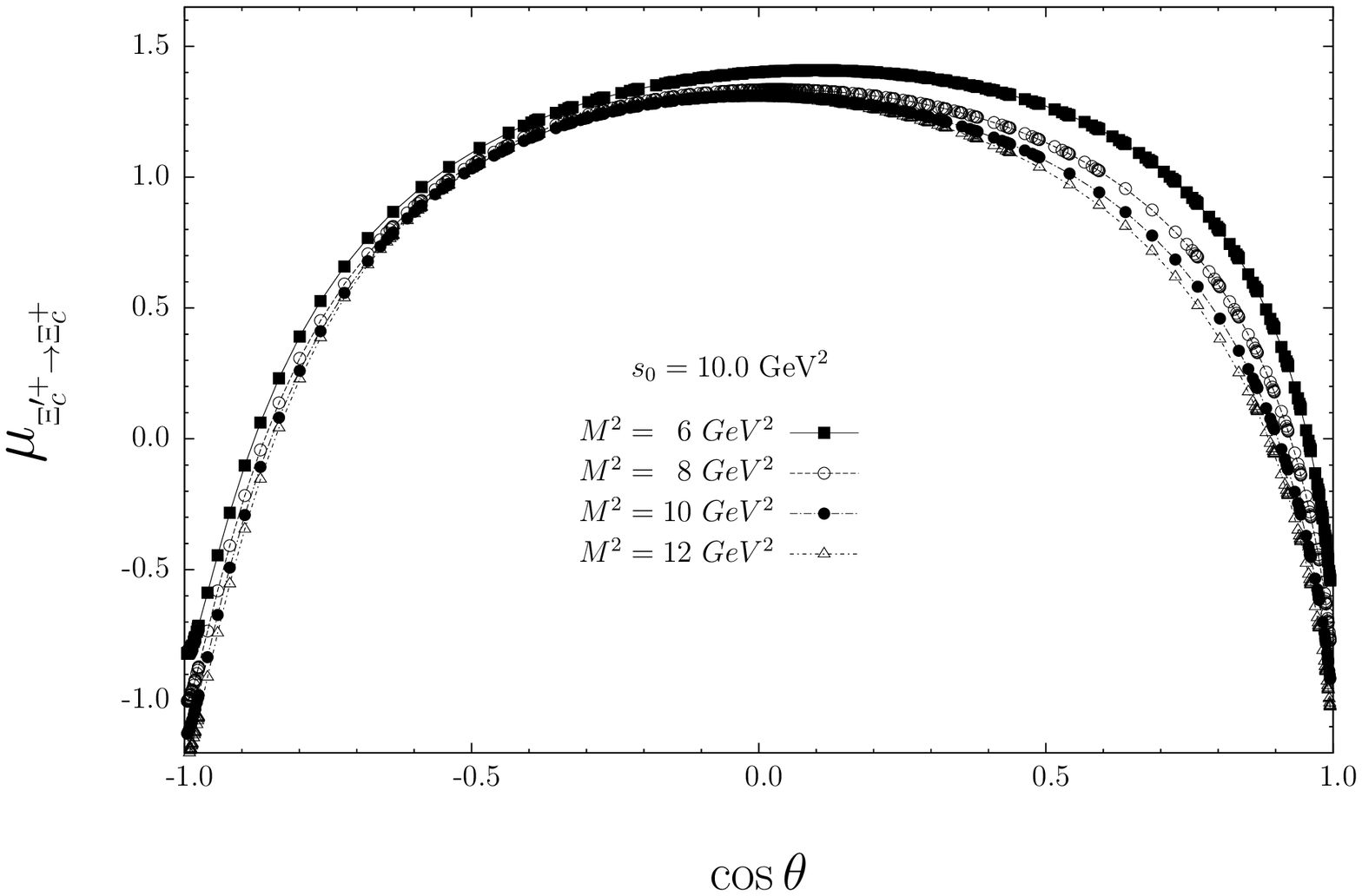}
\vskip 7.0cm   
\caption{}
\end{figure}

\begin{figure}
\vskip 3. cm
    \includegraphics{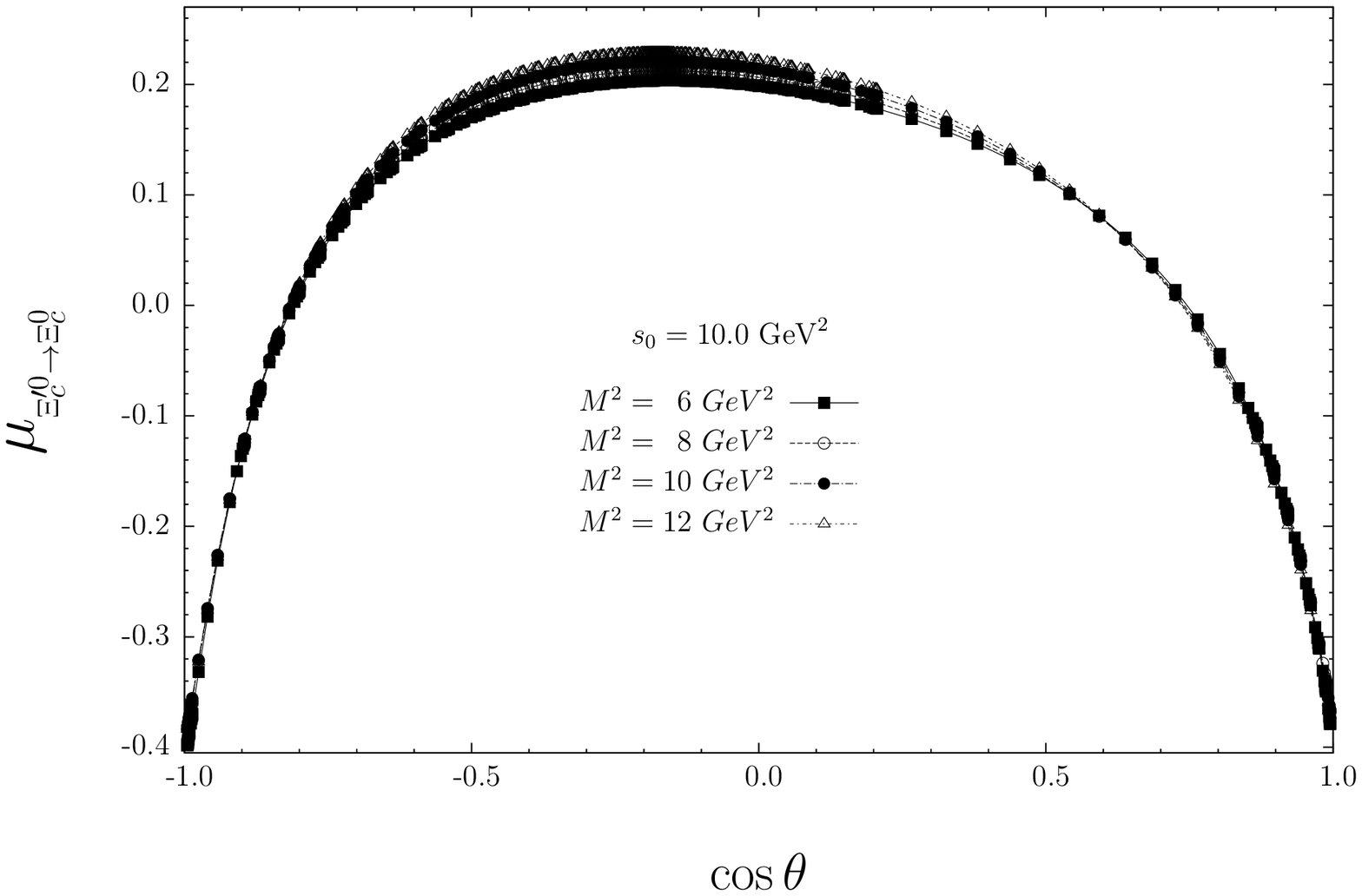}
\vskip 7.0cm
\caption{}
\end{figure}

\end{document}